\documentclass[twocolumn]{aastex63}

\usepackage{amsmath}

\received{August 10, 2021}
\accepted{November 14, 2021}
\submitjournal{ApJ Letters}

\begin{document}
\title{Critical Stellar Central Densities Drive Galaxy Quenching in the Nearby Universe}

\correspondingauthor{Bingxiao Xu, Yingjie Peng}
\email{bxu6@pku.edu.cn, yjpeng@pku.edu.cn}

\author{Bingxiao Xu}
\affiliation{Kavli Institute for Astronomy and Astrophysics, Peking University, 5 Yiheyuan Road, Beijing, 100871, P. R. China}

\author{Yingjie Peng}
\affiliation{Kavli Institute for Astronomy and Astrophysics, Peking University, 5 Yiheyuan Road, Beijing, 100871, P. R. China}

\begin{abstract}
		We study the structural and environmental dependence of the star 
		formation on the plane of stellar mass versus central core density
		($\Sigma_{\rm 1\ kpc}$) in the nearby universe. We study the central 
		galaxies in the sparse environment and find a characteristic 
		population-averaged $\rm \Sigma_{1\ kpc} \sim 10^9-10^{9.2}\ 
		M_{\odot}\ kpc^{-2}$, above which quenching is operating. This $\rm 
		\Sigma^{crit}_{1\ kpc}$ only weakly depends on the stellar mass, 
		suggesting that the mass-quenching of the central galaxies is more 
		closely related to the processes that operate in the central regions
        than over the entire galaxies. For satellites, at a given stellar mass, environment-quenching appears 
		to operate in a similar fashion as mass-quenching in centrals, also 
		starting from galaxies with high $\rm \Sigma_{1\ kpc}$ to low $\rm 
		\Sigma_{1\ kpc}$, and $\rm \Sigma^{crit}_{1\ kpc}$ becomes strongly 
		mass-dependent, in particular in dense regions. This is because
		(1) more low-mass satellites are quenched by the environmental 
		effects in denser regions and (2) at fixed stellar mass and 
		environment, the environment-quenched satellites have, on average,
		larger $\Sigma_{\rm 1\ kpc}$, $\rm M_{1\ kpc}/M_{\star}$ and 
		Sersic index $n$, and as well as smaller size. These results imply 
		that either some dynamical processes change the structure of the 
		satellites during quenching or the satellites with higher 
		$\Sigma_{\rm 1\ kpc}$ are more susceptible to the environmental effects. 
\end{abstract}


\section{Introduction} \label{sec:intro}
One of the core aims of galaxy evolution is to understand how galaxies shut 
off their star formation. The processes to quench the star formation can be 
broadly classified into two categories \citep{kau03b,bal06,pen10}: 
mass-quenching (internally driven processes, operating in both central and 
satellite galaxies) and environment-quenching (externally driven 
processes, operating in satellite galaxies). Active galactic nucleus (AGN) 
feedback has often been proposed as a plausible mass-quenching process in 
massive galaxies \citep{cro06,dar15,dar16,lin16,del19}. Other candidate 
mass-quenching 
processes include the morphological quenching \citep{mar09,gen14,gen20}, 
bar quenching \citep{gav15,kho18}, and angular momentum quenching 
\citep{pen20,ren20}. Meanwhile, in dense environments, the star formation is
mainly terminated by various environmental effects, such as ram-pressure 
stripping by the hot gas in the galaxy clusters \citep{gun72,aba99}, 
``strangulation" by cutting off the gas supply to the galaxies 
\citep{bal97}, tidal interaction \citep{sob11}, mergers \citep{pen10, 
pen12}, etc.

Interestingly, the quiescence of galaxies is found to be correlated with 
their structural parameters. It has been well established that the
quiescent galaxies (QGs) are generally smaller (or more compact) than 
star-forming galaxies (SFGs) at the same stellar mass, and in most cases,
QGs also possess prominent bulges. For instance, QGs have a higher surface
mass density within the effective radius $\rm \Sigma_e = 0.5M_{\star}/\pi 
R^2_e$ than SFGs at the same stellar mass in the local universe 
\citep{kau03a} and higher redshift \citep{fra08}. Separately, the Sersic 
indices of the local quenched galaxies are also found to be higher 
\citep{bel08,kau12,blu14}, and the same was shown for distant galaxies 
\citep{bel12,wuy12}. After testing numerous parameters, 
\citet{che12} found that the surface mass density within the inner 1 kpc 
region, $\Sigma_{\rm 1\ kpc}$, has higher power in discriminating the color 
than $\Sigma_e$ and is the best indicator of quenching at $z \sim 0.7$. 
Meanwhile, several other studies concluded that the central velocity dispersion within 
1 kpc, $\sigma_1$, is the best predictor of quiescence \citep{wak12,blu16,
tei16}, which is not entirely surprising, since $\sigma_1$ is shown to be 
well correlated with $\Sigma_{\rm 1\ kpc}$, at least for the massive 
galaxies with $\rm M_{\star} > 10^{10}M_{\odot}$ \citep{fan13}.

Since $\Sigma_{\rm 1\ kpc}$ is a sensitive indicator of quiescence, the distribution of galaxies with different star-forming levels on the $\rm M_{\star}-\Sigma_{1\ kpc}$ plane becomes a useful diagnostic to unveil the 
structural evolution of galaxies during quenching. \citet{fan13} studied a sample of nearby Sloan Digital Sky Survey (SDSS) 
galaxies and found that at fixed stellar mass, the specific star formation 
rate (sSFR) rapidly declines as $\Sigma_{\rm 1\ kpc}$ reaches a certain 
value, which implies a mass-dependent quenching threshold. They also
found that the quiescent population follows a tight scaling relation on the
$\rm M_{\star}-\Sigma_{\rm 1\ kpc}$ plane with a slope $\sim 0.64$. \citet{van14,tac15} and \citet{bar17} further extended the work to higher redshift and confirmed 
that such tight relation for QGs on the $\rm M_{\star}-\Sigma_{\rm 1\ kpc}$ plane has 
already been in place since $z\sim 2.5$. On the other hand, some SFGs likely
occupy the same position as the QGs do on the $\rm M_{\star}-\Sigma_{\rm 1\ 
kpc}$ plane, so the mass-dependent threshold or scaling relation for 
quiescence is only a necessary but not sufficient condition for quenching. 
Therefore, if we aim to predict the star-forming levels of an individual 
galaxy given its stellar mass and $\Sigma_{\rm 1\ kpc}$, a 
population-averaged indicator would be more informative, since the 
population-averaged star-forming levels contain the information of the 
relative abundance of the star-forming and quiescent populations at a given 
position on the $\rm M_{\star}-\Sigma_{\rm 1\ kpc}$ plane. Moreover, it is 
believed that in general, the environmental effects will not cause a 
significant morphological change in the galaxies. Most of the previous works
are restricted to exploring the central galaxies on the $\rm 
M_{\star}-\Sigma_{\rm 1\ kpc}$ plane, where internal processes might play more significant roles. However, recent 
studies show that the quiescence of satellite galaxies is also correlated 
with their morphology. For instance, the $\Sigma_{\rm 1\ kpc}$ of QGs is found 
to be systematically higher than that of SFGs at the same stellar mass 
\citep{kaw17,woo17,soc19,guo21}. These results call for a similar 
investigation on the distribution of star-forming levels on the $\rm M_{\star}-\Sigma_{\rm 1\ kpc}$ 
plane for satellite galaxies in order to shed light on the underlying physics of 
environment quenching. 

In this Letter, we utilize a sample of nearby SDSS galaxies to explore the 
population-averaged star-forming levels on the $\rm M_{\star}-\Sigma_{1\ kpc}$
plane, which is equivalent to the likelihood of quiescence of a galaxy 
given its position on the plane. We divided the sample into central and 
satellite galaxies and compute the local galaxy density to characterize the 
environment of the galaxies. The stellar mass range of our sample is down to 
$\rm log(M_{\star}/M_{\odot}) = 9$, which is sufficient to address the 
environmental impact on the quenching and morphology of the 
low-mass galaxies. Throughout, we adopt the following cosmological 
parameters where appropriate: $H_0$ = 70 km $s^{-1}$Mpc$^{-1}$, $\Omega_m$ =
0.3, and $\Omega_{\lambda}$ = 0.7.

\section{The Data} \label{sec:data}

\subsection{SDSS Sample}
The main galaxy sample used in this paper is the same sample used in 
\citet{pen10}, which was constructed from the parent 
SDSS DR7 catalog \citep{aba09}. The redshift range is $0.02 < z < 0.085$, which guarantees reliable spectroscopic redshift measurements. Each 
galaxy is weighted by 1/TSR $\times$ 1/$V_{\rm max}$, where TSR is the spatial
target sampling rate, determined using the fraction of objects that have 
spectra in the parent photometric sample within the minimum SDSS fiber 
spacing of 55" of a given object. The $V_{\rm max}$ values are derived
from the $k$-correction program version 4.2 \citep{bla07}. The use 
of $V_{\rm max}$ weighting allows us to include representatives of the 
galaxy population down to a stellar mass of about $10^9M_{\odot}$.

\subsection{Central 1\ kpc Surface Mass Density}

We compute the central 1 kpc surface mass density $\Sigma_{\rm 1\ 
kpc}$ by directly integrating the Sersic light profile and scaling the 
integrated luminosity within the inner 1 kpc. This method has been widely 
used in many previous studies \citep{bez09,whi17,kaw17} and is described as 
follows. The two-dimensional Sersic light profile can be described in the 
form of
\begin{equation} \label{eqn:sersic}
        I(r) = I_0 {\rm exp}\left[-b_n\left(\frac{r}{r_e}\right)^{1/n}\right],
\end{equation}
where $I_0$ is the central intensity, $n$ is the Sersic indices, $r_{\rm 
eff}$ is the circularized effective radii, and $b_n$ is defined as \citep{cio99}:
\begin{equation}
    b_n \approx 2n - \frac{1}{3} + \frac{4}{405n} + \frac{46}{25515n^2}.
\end{equation}
For the disk galaxies with Sersic indices $n < 2.5$ \citep{ken15}, the total
luminosity is obtained by integrating over the two-dimensional light profile
(Equation \ref{eqn:sersic}). We then convert the total luminosity to the 
total stellar mass, assuming that the mass follows the light and that there
are no strong color gradients. Finally, we calculate the stellar mass 
surface density in the inner 1 kpc by numerically integrating the following 
equation:
\begin{equation}
\Sigma_{\rm 1\ kpc} = \frac{\int_0^{\rm 1\ kpc}I(r)rdr }{\int_0^{\infty}I(r)rdr}    \frac{M_{\star}}{\pi(\rm 1\ kpc)^2}, \quad  n < 2.5
\end{equation}
where $\rm M_{\star}$ is the total stellar mass of the galaxy from the 
MPA/JHU DR7 value-added catalog. For the galaxies with prominent bulge components 
with $n > 2.5$, we assume that they follow spherical light profiles and 
perform an Abel transform to deproject the circularized, three-dimensional 
light profile \citep{bez09}:
\begin{equation}  \label{eqn:abel}
\rho\left(\frac{r}{r_e}\right) 
		 =  \frac{b_n}{\pi}\frac{I_0}{r_e}
             \left(\frac{r}{r_e}\right)^{1/n-1} \times \int_1^{\infty} \frac{{\rm exp}[-b_n(r/r_e)^{1/n}t]}{\sqrt{t^{2n} - 1}}dt.
\end{equation}
The total luminosity in this case is derived by integrating over the above
three-dimensional light profile, and the central surface mass density is 
given as
\begin{equation}
  \Sigma_{\rm 1\ kpc} = \frac{\int_0^{\rm 1\ kpc}\rho(r)r^2dr }{\int_0^{\infty}\rho(r)r^2dr}\frac{M_{\star}}{\pi(\rm 1\ kpc)^2}, \quad  n > 2.5. 
\end{equation}

\subsection{Local Environmental Indicators of Galaxies} \label{sec:massseg}
We characterize the environment of galaxies by their projected local 
overdensity. We estimate the local overdensity using the distance to the $N$th
nearest neighbor, where $N = 5$ in this study. Then, the dimensionless 
overdensity $1+\delta$ is defined as \citep{pen12} 
\begin{equation}
(1 + \delta)_5 = 1 + \frac{\Sigma_5 - \langle\Sigma\rangle}
             {\langle\Sigma\rangle},
\end{equation}		

The overdensity is computed from the volume of the cylinder that centered on 
each galaxy with a length $\pm$1000 $\rm km s^{-1}$. All the five closest 
neighbor galaxies have $M_{B, AB} \le -19.3 - z$, where $-z$ is used to
approximately account for the luminosity evolution of both passive and 
active galaxies.

The group catalog that we use in this work is an SDSS DR7 group catalog 
constructed with the technique outlined in \citet{yan07}. All galaxies 
are classified as either central galaxies or satellite galaxies. We required
the central galaxies be simultaneously both the most massive and the most
luminous (in the r band) galaxies within a given group. The group catalogs are
then cross-matched with our main galaxy sample.

\subsection{Photometry and Physical Properties}
Integrated photometries in six bands were used in this study: the near-UV (NUV) from the Galaxy Evolution Explorer ($GALEX$) and $ugriz$ from SDSS. The photometries were corrected for 
Galactic extinction and $k$-weighted to $z = 0$ using version 4.2 of the $k$-correct
code package described in \citet{bla07}. The spectroscopic
redshifts, total stellar mass, fiber velocity dispersion, and median
signal-to-noise ratios (S/Ns) in the spectra were obtained from the MPA/JHU
DR7 value-added catalog.\footnote{http://www.mpa-garching.mpg.de/SDSS/DR7/} The 
stellar masses were computed by fitting the integrated SDSS photometry with 
the stellar population models (similar in spirit to the method used in 
\citet{sal07}). We also extract morphological parameters such as the 
effective radius $R_e$, galaxy Sersic indices $n$, and ellipticity $e$ from 
the \citet{sim11} catalog of bulge+disk photometric decompositions. The axis ratio is computed as 
$b/a = 1 - e$ as defined.

We keep galaxies above the SDSS spectroscopic limit (r = 17.77) and the GALEX 
magnitude limit (NUV = 23) and with the stellar mass $\rm log(M/M_{\odot}) > 9$. 
In addition, a galaxy with a low axis ratio is more affected by the effect of
dust extinction, which would introduce error in measuring the size, Sersic 
indices, and $\Sigma_{\rm 1\ kpc}$, so an axis ratio cut $b/a > 0.5$ is 
applied to minimize the effects of dust extinction. We discard 56,634 
galaxies with low axis ratio, and a final sample of 89,469 galaxies makes the cut.

\section{The structural and environmental impact on quenching}
\subsection{The $\rm (NUV - r)$ Color and $\Sigma_{\rm 1\ kpc}$}
\begin{figure*}
   \plotone{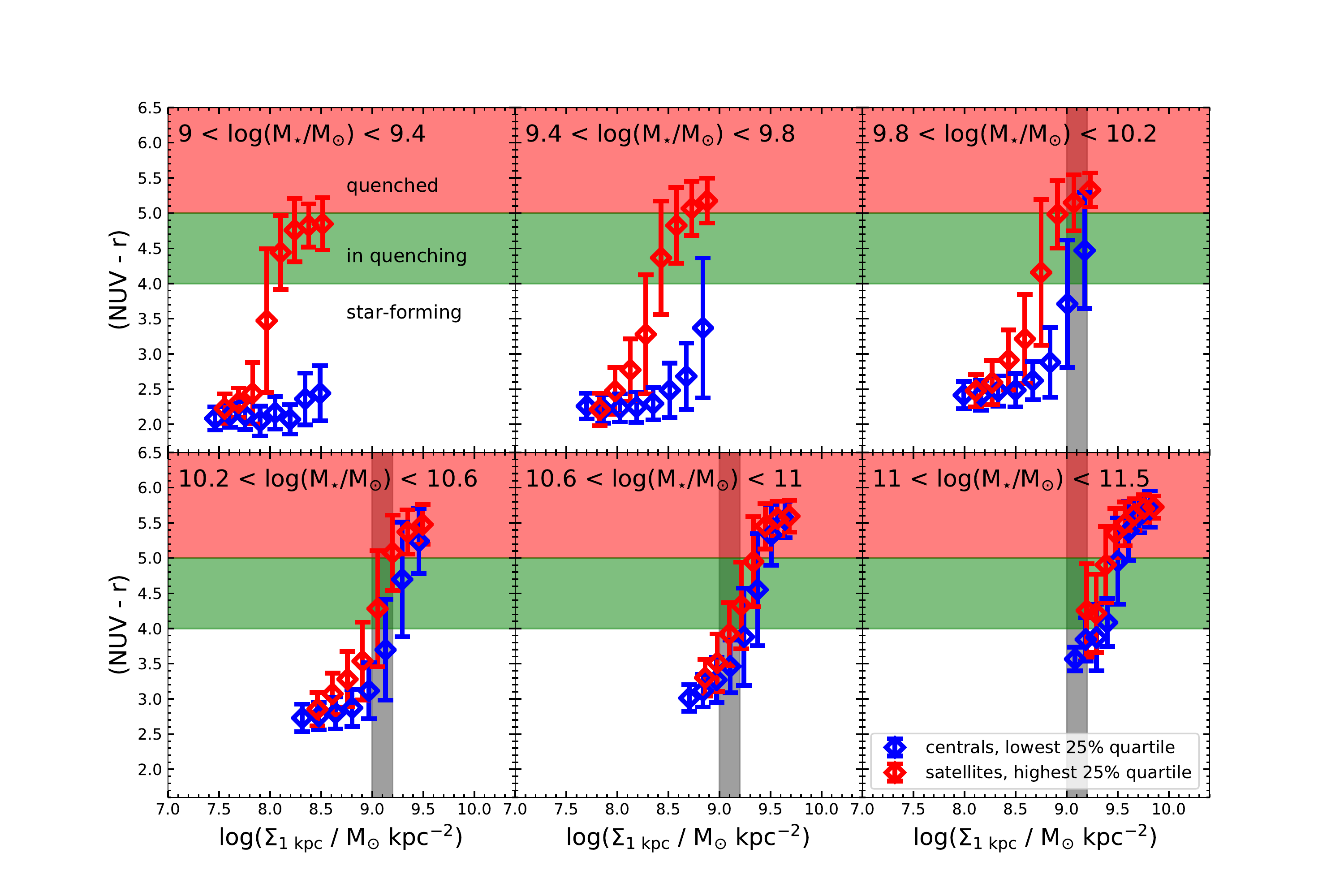}
		\caption{$V_{\rm max}$-weighted median (NUV - r) color as a 
		function of $\Sigma_{\rm 1\ kpc}$ in six stellar mass bins. The blue 
		(red) diamonds represent the central (satellite) galaxies in the 
		lowest (highest) 25\% density quartile. The green (red) shading marks
		the region with $\rm 4 < NUV - r < 5$ ($\rm NUV - r > 5$). The gray 
		strip denotes the transitional $\Sigma_{\rm 1\ kpc}$ of the central 
		galaxies when they first enter the green region, which is $\rm \sim 
        10^9-10^{9.2}\ M_{\odot}\ kpc^{-2}$ \label{fig:scatter} }

\end{figure*}

In Figure \ref{fig:scatter} we show the median (NUV - r) color as a function
of $\Sigma_{\rm 1\ kpc}$ in six stellar mass bins for central and satellite
galaxies in two extreme environments. The color in each bin is $V_{\rm 
max}$-weighted to account for the incompleteness for low-mass galaxies. Each
plot is divided into three regions based on their color as the star-forming 
indicator: a star-forming region with $\rm (NUV - r) < 4$, in quenching 
with $\rm 4 < (NUV - r) < 5$ and a quenched region with $\rm (NUV - r) 
> 5$ \citep{fan13}. For low-mass central galaxies (blue diamonds) with $\rm 
log(M/M_{\odot}) < 10$ in the sparsest environment, the median color remains
blue but starts to increase at high $\Sigma_{\rm 1\ kpc}$ in the bin of $\rm 
9.4 < log(M/M_{\odot}) < 9.8$. It indicates that the probability for 
low-mass central galaxies to be quenched is generally low, which is consistent with that
the internal quenching process being mainly determined by their stellar mass. As 
the stellar mass grows to $\rm log(M/M_{\odot}) > 10$, the central galaxies 
with higher $\Sigma_{\rm 1\ kpc}$ enter the green region and ignite the 
main process of quenching. Interestingly, the critical $\Sigma_{\rm 1\ kpc}$ that marks the 
quenching in process (the gray strip) appears to maintain at $\sim 10^9-
10^{9.2}$ $\rm M_{\odot}\ kpc^{-2}$ over a broad range of stellar mass from 
$10^{9.8}-10^{11.5}$ $\rm M_{\odot}$. On the other hand, the satellite galaxies 
in the densest environment (red diamonds) with high $\Sigma_{\rm 1\ kpc}$ are 
quenched in all mass bins. The critical $\Sigma_{\rm 1\ kpc}$ for satellites is 
typically smaller than that of central galaxies at fixed stellar mass, and the 
difference in $\rm \Sigma^{crit}_{1\ kpc}$ becomes larger as the stellar mass 
decreases and almost vanishes when $\rm log(M/M_{\odot}) > 10.6$. This implies 
that the environmental quenching is most effective for low-mass satellites and the 
internal mass-quenching processes dominate in massive galaxies, which is in 
broad consistency with previous studies.

\subsection{The Dependence of Color on the Stellar Mass and $\Sigma_{\rm 1\ kpc}$} \label{sec:ms+sigma1}
\begin{figure*}
   \plotone{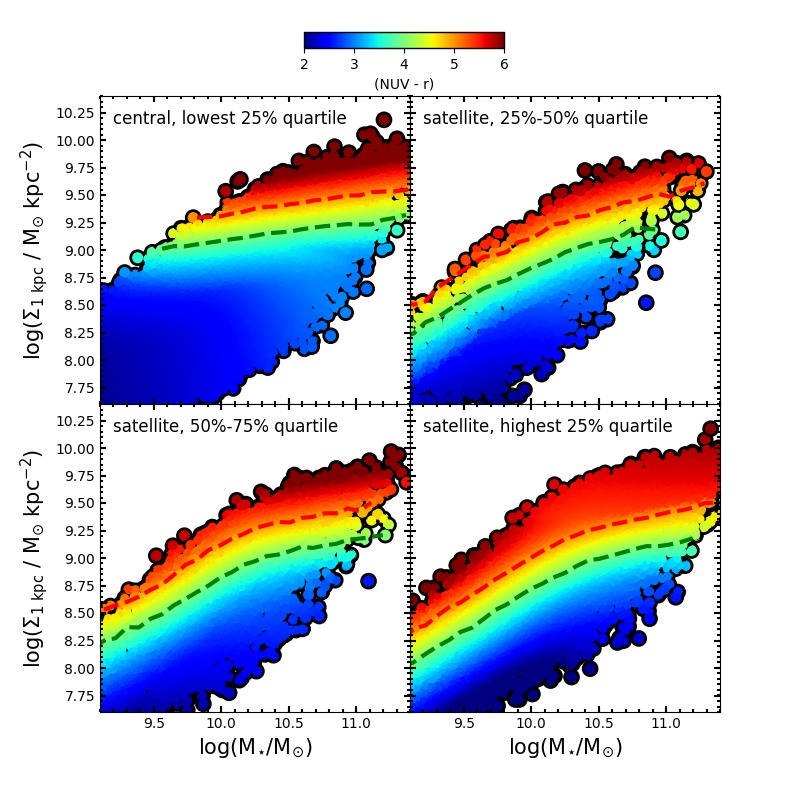}
        \caption{Central 1\ kpc surface mass density $\Sigma_{\rm 1\ kpc}$
        as a function of stellar mass in four environmental bins with
        increasing log($1+\delta$), color-coded by the (NUV - r) color. The
		data have been $V_{\rm max}$-weighted and LOESS-smoothed. The green 
    	and red dashed lines denote the transitional $\Sigma_{\rm 1\ kpc}$ in 
	    quenching with (NUV - r) $\sim$ 4 and in quenched status with (NUV 
		- r) $\sim$ 5, respectively. \label{fig:compar} }

\end{figure*}

To take a closer look at the different dependence of $\Sigma^{\rm crit}_{\rm
1\ kpc}$ on the stellar mass for the centrals and satellites shown in Figure
\ref{fig:scatter}, in Figure \ref{fig:compar} we present the central surface
mass density $\Sigma_{\rm 1\ kpc}$ as a function of stellar mass $\rm 
M_{\star}$ color-coded by the median (NUV - r) color. To reveal their 
environmental dependence, we further assign the galaxies to four 
environment bins. The galaxies selected in the leftmost panel are 
the central galaxies in the lowest overdensity $\rm log(1+\delta)$ 
quartile. Such selection is to maximize the purity of the central galaxies, 
since no group finder operates perfectly in classifying the centrals and 
satellites, and any ``over-fragmentation" or ``over-merging" of groups will 
lead to misclassification of satellites as centrals, and vice versa 
\citep{pen12}. We ignore the central galaxies in denser environments, since 
the color distribution is insensitive to the change of overdensity for 
centrals (see Appendix \ref{appen:central}) and only take account of the 
satellites with increasing $\rm log(1+\delta)$ in the rest of the panels. For each data point, we perform a
$V_{\rm max}$ weighting correction inside a box of 0.3 $\times$ 0.2 dex$^2$ 
that centers on each data point. To better visualize the trend, we further 
smooth the data using the locally weighted regression method LOESS 
\citep{cle88} as implemented by \citet{cap13}. LOESS is extremely useful in 
estimating the average values in bins whose bin size is small and unveiling
the overall underlying trends by reducing the intrinsic and observational 
errors. We present the LOESS-smoothed version in Figure \ref{fig:compar}.

The stellar mass dependence of $\Sigma_{\rm 1\ kpc}$ during the course of 
quenching in different environments is displayed in terms of color 
distribution in Figure \ref{fig:compar}. To quantitatively catch the 
trend in $\Sigma_{\rm 1\ kpc}$ transitioning from star-forming to 
quenched status, in each plot, we divided the data into 25 stellar mass bins,
and define $\Sigma^{\rm crit}_{\rm 1\ kpc}$ in quenching (quenched) as the 
median $\Sigma_{\rm 1\ kpc}$ for galaxies that have $\rm 4(5) - 0.15 < (NUV - 
r) < 4(5) + 0.15$\footnote{The quadratic error for (NUV - r) is $\sim$ 0.8 
dex} in each stellar mass bin. Then we overplot the transition lines for 
in quenching (quenched) as green (red) dashed lines in Figure 
\ref{fig:compar} for reference.

For central galaxies in sparse environments, massive galaxies with 
higher $\Sigma_{\rm 1\ kpc}$ have a higher chance of shutting off their star formation.
For the transition from star-forming to quenching in operation, 
$\Sigma^{\rm crit}_{\rm 1\ kpc}$ only increases for $\sim$ 0.25 dex as the stellar
mass increases for $\sim$ 2 dex. Similarly, $\Sigma^{\rm crit}_{\rm 1\ kpc}$
for the transition from quenching in operation to quenched status is also
insensitive to the stellar mass. The weak dependence of $\Sigma^{\rm 
crit}_{\rm 1\ kpc}$ on the stellar mass suggests that the internal quenching
processes are more related to the processes that take place in the inner 
region of galaxies, rather than over entire galaxies.

The $\rm \Sigma^{crit}_{1\ kpc}$ for satellite galaxies exhibits more 
versatility; the massive satellites have very similar trends in $\rm 
\Sigma^{crit}_{1\ kpc}$ compared with their central counterparts. Such 
similarity may reflect similar quenching processes in massive centrals and 
satellites. It has been shown in previous studies that the internal quenching processes
operate in both massive centrals and satellites \citep{bal06,pen10}.
It is also consistent with the result in \citet{blu20b} which concluded that the quenching in massive 
satellites is largely determined by an internal parameter $\sigma_1$ rather 
than environmental effects. On the other hand, $\rm \Sigma^{crit}_{1\ 
kpc}$ for low-mass satellites becomes strongly stellar mass-dependent as the
overdensity increases. For instance, the difference in $\rm 
\Sigma^{crit}_{1\ kpc}$ between the centrals and satellites at $\rm 
log(M/M_{\odot})$ = 9.5 and 9.8 is $\sim$ 0.4 and 0.25 dex in the $25\%-50\%$ 
quartile and increases to 0.75 and 0.5 dex in the highest quartile. It 
demonstrates that the environmental effects are more prominent in low-mass 
galaxies. Surprisingly, not all satellites are quenched, even in the densest 
environment. Instead, at a fixed mass, those satellites with higher 
$\Sigma_{\rm 1\ kpc}$ are more likely to be quenched, which can be 
identified by the stratification in their color distribution. Moreover, $\rm
\Sigma^{crit}_{1\ kpc}$ appears to be lower as log($1+\delta$) increases at 
fixed mass, which is manifested by the gradual ``bending" of the curves 
toward the low-mass end. We have also explored $\Sigma_{\rm 1\ kpc}$ as a 
function of stellar mass color-coded by other star-forming indicators, 
such as $\Delta$MS (the offset distance to the star-forming main-sequence 
line); all the trends remain similar. We will discuss these trends in detail
in Section \ref{sec:dis}.

\subsection{The Role of AGNs}
\begin{figure*}
     \plotone{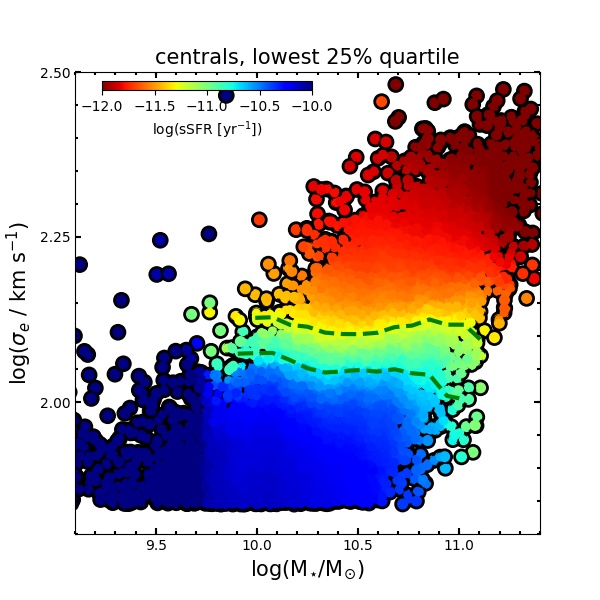}
     \caption{Velocity dispersion $\sigma_e$ as a function of stellar mass
       for the central galaxies in the lowest 25\% density quartile, color-coded
       by the LOESS-smoothed sSFR. The two green dashed lines denote the median
        $\sigma_e$ at sSFR $\sim$ $\rm 10^{-11 - 0.25} yr^{-1}$ and
        $\rm 10^{-11 + 0.25}yr^{-1}$, respectively.
                \label{fig:bhmass}}

\end{figure*}

The curve of $\Sigma^{\rm crit}_{\rm 1\ kpc}$ for the central galaxies in 
underdense regions is much flatter than the quenching ``boundaries" reported in previous 
studies \citep{fan13,che20}. This is not unexpected, since there are more 
SFGs at $\rm 9.5 < log(M/M_{\odot}) < 10$ to leverage the slope; more 
detailed environmental characterization (e.g. by the overdensity) could be 
another reason. We emphasize that this transitional 
curve is only true in a statistical sense, and it does not necessarily imply
any shortage of QGs below this line, but that the probability of an individual galaxy shutting down its star formation at a given 
$\Sigma_{\rm 1\ kpc}$ below the line is low. The weak mass dependence
appears to suggest a potential linkage between the mass-quenching processes 
and the physical processes that operate in the central regions of the galaxies. If this 
logic is correct, then a natural candidate for the quenching engine is the 
supermassive black hole \citep[SMBH;][]{cro08,che20}, since the SMBHs are
more closely related to the properties of the galactic bulges than the 
outskirts of the galaxies \citep{kor13}. The well-known $\rm M_{BH}-\sigma_e$ relation
\citep{mcc13,sag16} allows us to obtain a similar transitional curve of the 
black hole (BH) mass $\rm M_{BH}$ as well. To guarantee reliable
measurements of $\rm \sigma_e$, we apply a quality cut of $\rm S/N > 10$ to the spectra and only keep values of $\rm \sigma_{fib} > 70 km s^{-1}$,  
which is the instrumental resolution of the SDSS spectra. We then correct 
the fiber velocity dispersion $\rm \sigma_{fib}$ to the velocity dispersion within a one-eighth effective radius
as follows \citep{ber03}:
\begin{equation}
 \sigma_e = \rm \sigma_{fib}\left(\frac{r_{fib}}{r_0/8} \right)^{0.04},
\end{equation}
where $\rm r_{fib} = 1".5$ and $r_0$ is the circular effective radius
in arcseconds. We first assume the quenching is in process at $\rm sSFR^{crit}
\sim -11$ with a 0.25 dex variation,\footnote{Instead of (NUV - r) color, sSFR
is chosen to compare with other studies for convenience.} and the transitional
$\sigma^{\rm crit}_e$ is then defined as the median $\sigma_e$ that corresponds
to the upper and lower bounds of $\rm sSFR^{crit}$. The $V_{\rm 
max}$-weighted correction is computed inside a box of 0.3 $\times$ 0.2 dex$^2$ that centers on each data point, similar to what we did in Section \ref{sec:ms+sigma1}. We plot $\sigma_e$ as
a function of the stellar mass color-coded by sSFR for the central galaxies in underdense regions in Figure \ref{fig:bhmass}. The green dashed lines denote the
upper and lower bounds of $\rm sSFR^{crit}$. The transitional $\sigma^{\rm 
crit}_e$ shows a similar mass independence as that of $\Sigma^{\rm crit}_{\rm 
1\ kpc}$, which is expected, since there is a strong correlation 
between $\Sigma_{\rm 1\ kpc}$ and $\sigma_e$ \citep{fan13}. We then use the 
best-fit $\rm M_{BH}-\sigma_e$ relation in \citep{mcc13},

\begin{equation}
  \rm log(M_{BH}) = 5.20 \times log(\sigma_e / 200) + 8.39,
\end{equation}

to translate the $\sigma^{\rm crit}_e$ to $\rm M^{crit}_{BH}$, which is
$\sim 10^7-10^{7.4} M_{\odot}$.

We note that such a weakly stellar mass-dependent characteristic BH mass 
is also implemented in IllustrisTNG  \citep{ter20,zin20}, which 
successfully reproduces a wide range of observation properties of both 
star-forming and QG populations, such as the stellar mass 
functions and various scaling relations. In IllustrisTNG, once the BH of the galaxy exceeds a threshold mass at $\rm \sim 10^{8.2}\ 
M_{\odot}$, the kinetic mode of the AGN feedback is turned on. The isotropic
kinetic winds that are driven by the BH \citep{wei17,yua18} then 
effectively remove the gas and produce quiescence of the galaxy 
\citep{ter20,zin20}. This implementation of AGN feedback based on a 
threshold BH mass in TNG is in qualitative agreement with our observation result, though our $\rm
M^{crit}_{BH}$ is lower than their implemented value of $\rm \sim 10^{8.2}\ 
M_{\odot}$. An alternative possibility that relates the central mass density
to the quiescence is the morphological quenching \citep{mar09}, which 
proposes that the existence of central massive bulges tends to stabilize the
entire gas disks and prevent them from fragmenting into molecular clouds to 
continue the star formation. Nonetheless, it is possible that the overall 
effects of internal quenching could be many effects working in concert.

\subsection{The Structural Dependence of Satellite Quenching}
\begin{figure*}
        \plottwo{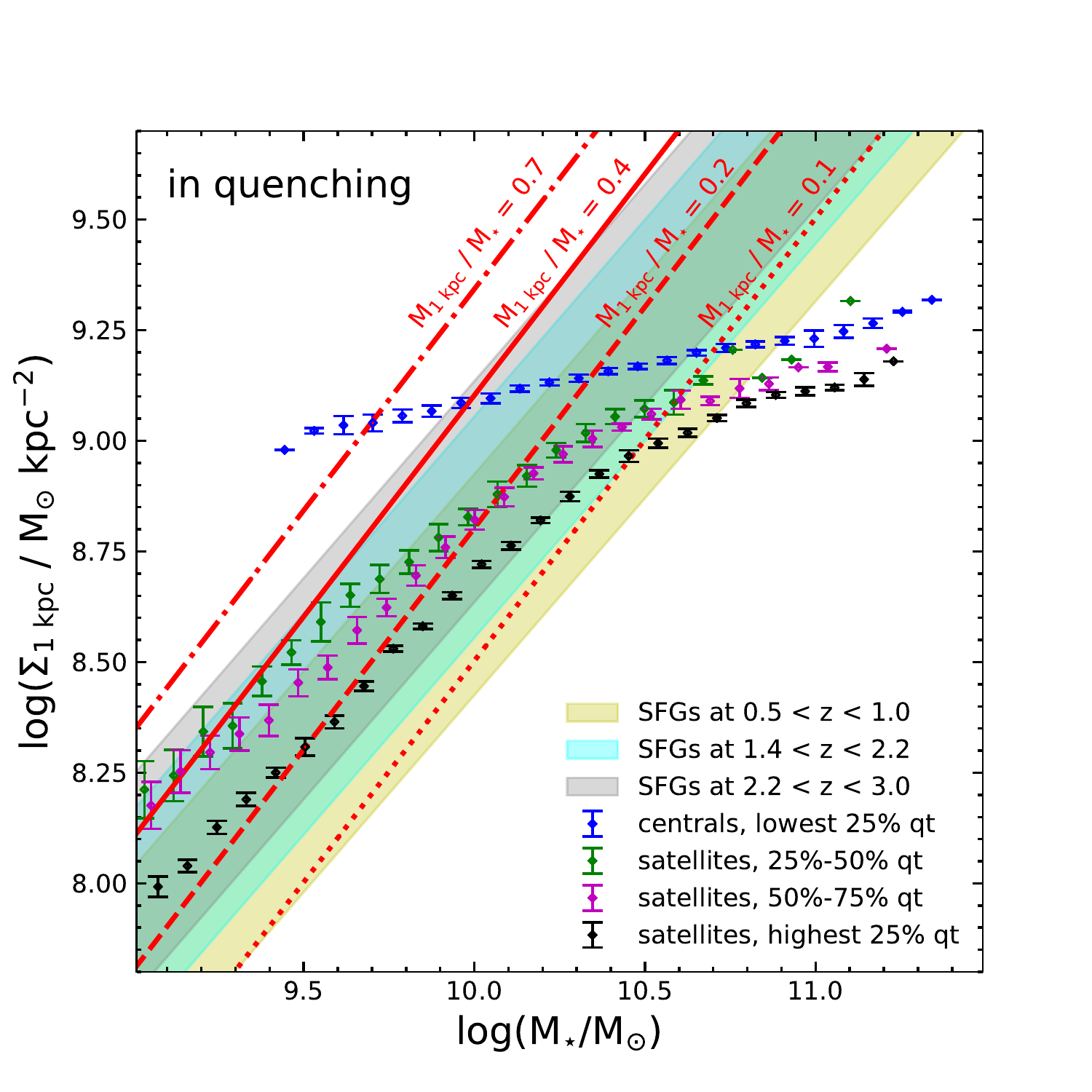}{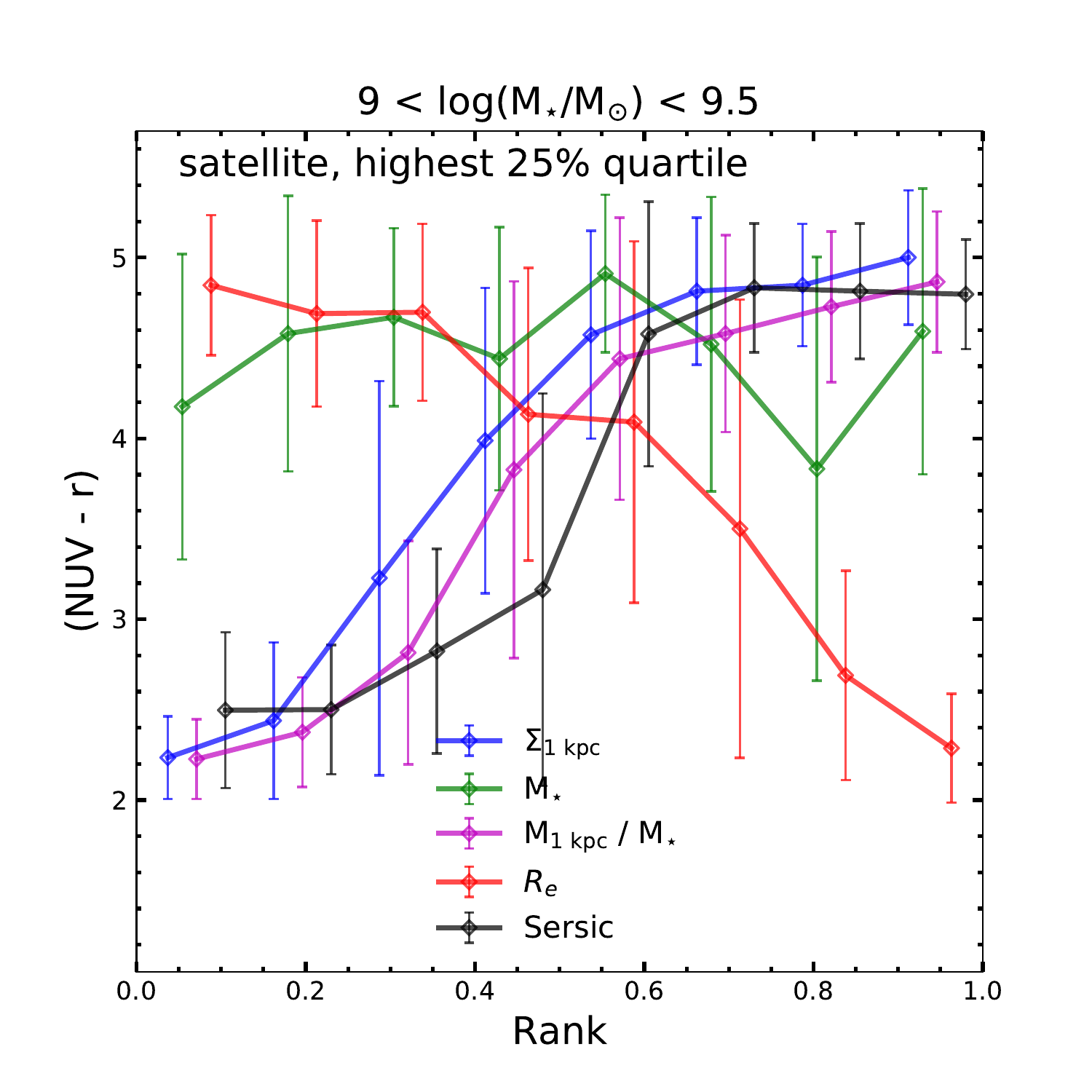}
	\caption{The left panel shows the transitional $\rm \Sigma^{crit}_{1\ kpc}$ in
	quenching (NUV- r $\sim$ 4) as a function of stellar mass. Different 
	colored dots denote galaxies in different environments. The error bar 
	denotes the 1$\sigma$ error in each mass bin. The dotted, dashed, solid, 
	and dash-dotted red lines mark the mass fraction of the central 1\ kpc 
	$\rm M_{1\ kpc}/M_{\star}$ = 0.1, 0.2, 0.4, and 0.7. The gray, cyan and 
	yellow shaded regions mark the average $M_{\star}-\Sigma_{\rm 1\ kpc}$ 
	relations for SFGs at $2.2 < z < 3.0$, $1.4 < z < 2.2$, $0.5 < z < 1.0$
	with $\pm 1\sigma$ scatter from \citet{bar17}. The right panel shows the
	$V_{\rm max}$-weighted median (NUV - r) color for satellite galaxies 
	at a given stellar mass $\rm 9 < log(M/M_{\odot}) < 9.5$ and a given 
	environment (the highest 25\% quartile) as a function of the rank of 
	$\Sigma_{\rm 1\ kpc}$, $M_{\star}$, $\rm M_{1\ kpc}/M_{\star}$, $R_e$ and 
	Sersic index $n$, respectively. 
		\label{fig:gthr} }
\end{figure*}

We further highlight the trends discussed in Section \ref{sec:ms+sigma1} in the 
left panel of Figure \ref{fig:gthr}. We directly plot $\rm \Sigma^{crit}_{1\ kpc}$
for the transition from star-forming to quenching in process\footnote{We only 
show the data of $\rm \Sigma^{crit}_{1\ kpc}$ for quenching in process, 
since the $\rm \Sigma^{crit}_{1\ kpc}$ for quenched status are almost parallel and $\sim$
0.25 dex higher} as a function of stellar mass in four environment bins. To 
assess the uncertainty of the $\rm \Sigma^{crit}_{1\ kpc}$ in each stellar 
mass bin, we jackknife resample the original data by randomly selecting 
80\% of the data points, and compute the curve of $\rm \Sigma^{crit}_{1\ kpc}$ in 
each environment bin 30 times, and compute the median value and standard 
deviation of the resampling data in each stellar mass bin. To quantitatively
describe the trend in $\rm \Sigma^{crit}_{1\ kpc}$, we parameterize $\rm 
\Sigma^{crit}_{1\ kpc}$ as a function of the stellar mass and overdensity 
in the form as follows (see details in Appendix \ref{appen:param}):
\begin{equation}\label{eqn:form}
		\rm \Sigma^{crit}_{1\ kpc}  =	\Sigma^0_{1\ kpc}10^{ln(1 - exp(-(M_{\star}/10^{10}M_{\odot})^{\alpha}))}, 
\end{equation}		
where $\rm \Sigma^0_{1\ kpc}$ is the normalization and $\alpha$ is the slope
of the power law at the low-mass end. The best-fit parameters are listed as 
follows (also see Figure \ref{fig:param+fit} in Appendix \ref{appen:param}):
\begin{widetext}
\begin{eqnarray}
				\rm for\ centrals \ (lowest-density quartile), \quad & \alpha = 0.1275 \pm 0.0030, & \nonumber\\
  \rm                \quad & \rm log\Sigma^0_{1\ kpc} = 9.5459 \pm 0.0023, & \nonumber\\
				\rm	 for\ satellites, \quad & \rm \alpha = (0.0891 \pm 0.0061) \times log(1+\delta) + (0.3454 \pm 0.0080), & \nonumber\\
  \rm               \quad & \rm  log\Sigma^0_{1\ kpc} = (-0.1175 \pm 0.0029) 
		\times log(1+\delta) + (9.3540 \pm 0.0039). & \label{eqn:fit}
\end{eqnarray}		
\end{widetext}
From the best-fit values, the slope and normalization of centrals
are systematically lower and higher than those of satellites; whereas for 
satellites, the slope increases with the overdensity and the normalization 
decreases with the overdensity (see Figure \ref{fig:param+oden} in Appendix
\ref{appen:param}). In addition, we overplot four lines that represent 
different values of the mass fraction of the central 1\ kpc $\rm M_{1\ kpc} / 
M_{\star}$ for comparison. Interestingly, the environmental effects operate
on the low-mass satellites in a way that $\Sigma^{\rm crit}_{\rm 1\ kpc}$ 
appears to be parallel to the iso[$\rm M_{1\ kpc}/M_{\star}$] lines as 
log($1+\delta$) increases, which is unexpected. For example, in the densest 
environment bin (black dots), the quenching in low-mass satellites is 
operating when $\rm M_{1\ kpc}/M_{\star} > 0.2$, regardless of their stellar
mass. We further study the sensitivity of color on different physical 
parameters for the satellite galaxies at $\rm 9 < log(M/M_{\odot}) < 9.5$ in
the dense environment. The mass range is selected to guarantee the dominance
of environmental impacts, since there are few quenched central galaxies with 
$\rm \Sigma_{1\ kpc} > \Sigma^{\rm crit}_{1\ kpc}$ in this mass range (see 
left panel of Figure \ref{fig:gthr}). The $V_{\rm max}$-weighted median 
color for each parameter is computed in an interval of 0.125 in the rank. 
The result is shown in the right panel of Figure \ref{fig:gthr}. In general,
the color of the satellites is sensitive to all of the structural parameters, such 
as $\Sigma_{\rm 1\ kpc}$, $\rm M_{1\ kpc}/M_{\star}$, $R_e$ and Sersic index
$n$, in the densest environment. The correlations for all of the structural 
parameters become noticeably flat as the color reaches above 5 when 
quenching is finished, which suggests that the morphology of the satellites is 
more closely related to their color before and in the main processes
of environment quenching. The variation in stellar mass has almost no 
impact on the median color of the satellite galaxies. We have also tested other 
stellar mass bins for satellites in the highest-density quartile and found a
similar strong dependence of the median color on $\Sigma_{\rm 1\ kpc}$, $\rm
M_{1\ kpc}/M_{\star}$,  $R_e$ and Sersic index $n$.

\section{discussion and summary}\label{sec:dis}
We utilize the SDSS DR7 data to study the structural dependence of star 
formation in nearby galaxies by investigating the distribution of their 
median (NUV-r) color on the $M_{\star}-\Sigma_{\rm 1\ kpc}$ plane. We 
separate the sample into central and satellite galaxies and use the local 
overdensity to characterize their environment. For the central galaxies in 
underdense regions where quenching is expected to be driven primarily by 
the internal mass-quenching process, we find that there exists a 
characteristic $\rm \Sigma^{crit}_{1\ kpc} \sim 10^9 -10^{9.2}\ M_{\odot}\ 
kpc^{-2}$ (Figure \ref{fig:scatter} \& \ref{fig:compar}) above which 
quenching is operating in a statistical sense. This $\rm \Sigma^{crit}_{1\ 
kpc}$ shows only a weak dependence on stellar mass (Figure \ref{fig:compar} and left panel of Figure
\ref{fig:gthr}), and can be roughly converted to a critical BH mass 
$\rm M^{crit}_{BH}$ of about $\rm 10^7M_{\odot}$. \citet{pio21} explored the
sSFR and quiescence fraction as a function of BH mass by using a similar 
SDSS sample of central galaxies. They found that the quenching starts to 
operate (sSFR drops below -11) when $\rm M_{BH} \sim 
10^{7}M_{\odot}$. \citet{blu16,blu20b} used galaxy samples from SDSS DR7 and
DR15 to study the relative importance of numerous parameters in driving the 
quenching process and found that the central velocity dispersion $\sigma_c$
is the most important factor for central galaxies and massive satellites. 
They further investigated the fraction of quenched spaxels $f_Q$ as a 
function of $\sigma_c$ and showed that the trends in $f_Q$ for centrals and 
satellites are indistinguishable when $\sigma_c > 100-120 \rm kms^{-1}$, 
below which the trends for the different populations begin to diverge. Such 
a critical $\sigma_c$ also corresponds to a BH mass of $\rm M_{BH} \sim 
10^{7}M_{\odot}$. Our result is fully consistent with their findings. The 
new result here is that we explore the $\rm \Sigma^{crit}_{1\ kpc}$ or $\rm 
M^{crit}_{BH}$ as a function of stellar mass, and we find that these 
critical values are only weakly dependent on the stellar mass for the 
central galaxies, which suggests that the mass-quenching of central galaxies
is closely related to the processes that are operative in the central 
regions of galaxies. Our result is qualitatively consistent with the AGN 
feedback model prescription in IllustrisTNG, though the $\rm M^{crit}_{BH} 
\sim 10^{8.2} M_{\odot}$ implemented in TNG is systematically higher than 
the inferred value in this work. The quantitative offset on the threshold BH
mass may suggest that more updates on the implementation of future numerical
simulations are needed to agree with our findings. Moreover, recent 
observations implemented by IFU have studied the spatially resolved star-forming profile for nearby galaxies and revealed that the quenching in massive 
central galaxies is likely to proceed via an ``inside-out" mode; these 
galaxies typically possess quenched cores and star-forming outskirts 
\citep{ell18,bel18,guo19,blu20b,zha21}. Our result suggests that such an 
``inside-out" quenching process should be accompanied by an increment in the
central mass density, which is likely related to the buildup of central 
bulges. This is also supported by \citet{zha21}, who found that the internal
quenching is closely related to the compaction and growth of the bulge 
component. However, it should be noted that the correlation does not 
necessarily lead to causality, and direction of the causal relation between 
the quenching and structural change is still uncertain at this stage.

Another new and interesting finding in this work is that, as shown in 
Figure \ref{fig:compar}, at any fixed stellar mass (except at the most 
massive end) and environment of satellite galaxies, the median color always 
increases rapidly with $\rm \Sigma_{1\ kpc}$.  In the right panel of Figure 
\ref{fig:gthr}, we also show the results of $\rm M_{1\ kpc}/M_{\star}$, 
Sersic index $n$ and $R_e$ for satellites with $\rm 9 < log(M_{\star}/
M_{\odot}) < 9.5$, and the trends in other mass and environment bins remain
similar. It should be noted that these trends are similar for both massive 
galaxies (centrals and satellites, where mass-quenching dominates) 
and low-mass satellite galaxies (where environment-quenching dominates).
Hence, it indicates that environment-quenching appears to operate in a 
similar fashion as mass-quenching, and both operate from the
galaxies with high $\rm \Sigma_{1\ kpc}$ to low $\rm \Sigma_{1\ kpc}$,  
although their underlying physical mechanisms could be very different 
(internal versus external origin).

Meanwhile, it is well established that at a fixed mass, satellite 
galaxies in a denser environment have a larger quiescence fraction (due to 
a stronger environmental effect), in particular for the low-mass satellites 
where environment-quenching dominates \citep{bal06,pen10}. As noted 
above, environment-quenching tends to produce quenched satellites with 
higher $\rm \Sigma_{1\ kpc}$ at a given stellar mass, which leads to the 
increasing red regions in the upper part of the galaxy distribution at the 
low-mass end, when the environmental effect becomes progressively stronger,
i.e. from the first panel to the last panel in Figure \ref{fig:compar}.  
However, there is little change in the color distribution at the massive 
end, as most of these massive galaxies were mass-quenched (i.e. via the 
internal quenching channel) in both dense and underdense environments. Hence, this produces a bent transitional $\rm \Sigma^{crit}_{1\ kpc}$ in 
quenching (as shown by the curves in Figure \ref{fig:compar} and left panel 
of Figure \ref{fig:gthr}), with a stronger bending in denser environments.

Following the fundamental formation relation (FFR) proposed in \citet{dou21a,
dou21b}, the molecular gas fraction of galaxies is tightly correlated with their
sSFR. At fixed stellar mass and environment, compact galaxies are, on 
average, redder and hence have lower sSFR and lower molecular gas fractions. Compaction 
during the environmental quenching process can be triggered by, for instance, 
major or minor mergers \citep{bar91,hop10} and tidal compression \citep{dek03}. 
However, the merger rate (which is observed to decrease with redshift, e.g. 
\citet{fer20}) is relatively low in the local universe, and the local compact 
satellites quenched by the environmental effects (with $\Sigma_{\rm 1\ kpc}$
below the $\rm \Sigma^{crit}_{1\ kpc}$ of centrals) cannot all be explained by the 
mergers that occurred in the local universe. The progenitors of the local quenched satellites 
should be the SFGs at redshift z $\sim$ 0.5 or higher 
\citep{pen15}, so the compaction or tidal effects may happen at higher 
redshift, where the merger activities are more frequent. Another possibility is 
that the structural dependence of satellite quenching is due to the ``progenitor bias", which does not require the structural change during quenching. The 
progenitors of the local quenched satellites are the SFGs at 
higher redshift, which are more compact at the same stellar mass, and the 
SFGs that fall into the halo at a later time have lower 
$\Sigma_{\rm 1\ kpc}$ and have yet to be quenched. We show the average $M_{\star}-
\Sigma_{\rm 1\ kpc}$ relations for SFGs \citep{bar17} at 
high redshifts to compare with the structure of the quenched satellites in the 
local universe in the left panel of Figure \ref{fig:gthr}. From $z > 2$ to $ < 
1$, the average $\Sigma_{\rm 1\ kpc}$ of SFGs decreases for 
$\sim$ 0.2 dex. Therefore, if we assume that $\Sigma_{\rm 1\ kpc}$ does not 
change during the process of environment-quenching, and that the progenitors
of the local quenched satellites are SFGs at $z > 2$, the 
change in $\Sigma_{\rm 1\ kpc}$ due to progenitor bias is then $\sim$ 0.2 dex.
However, this value is too small to explain the strong correlation between 
$\Sigma_{\rm 1\ kpc}$ and color shown in the right panel of Figure 
\ref{fig:gthr}. Hence, the role of the progenitor bias should be minor.

Alternatively, the strong structurally dependent environmental quenching can be 
explained by the fact that the satellite galaxies with higher $\Sigma_{\rm 1\ kpc}$ are 
more susceptible to the environmental effects and are more likely to be 
environment-quenched. It might be related to the ``preprocessing" that
a fraction of galaxies have been orbiting in smaller subhalos prior to their 
accretion onto the current host halo \citep{mcg09,hou14}, and the 
morphological transformation in satellites is likely to be a long-lasting 
process and has been operative since their star-forming stage. Future cold 
gas surveys in both H I and H$_2$ and comparison with the structure of the 
simulated quenched satellites will provide the critical data and insight 
needed for a more detailed investigation of this issue. \\

We gratefully acknowledge the anonymous referee for the comments and 
criticisms that have improved the paper. We acknowledge National Science 
Foundation of China (NSFC) grant Nos. 12125301, 11773001, 11721303, 11991052
and the science research grants from the China Manned Space Project No.
CMS-CSST-2021-A07.

\appendix
\section{The Critical central density for central galaxies} \label{appen:central}
\begin{figure*}
   \plotone{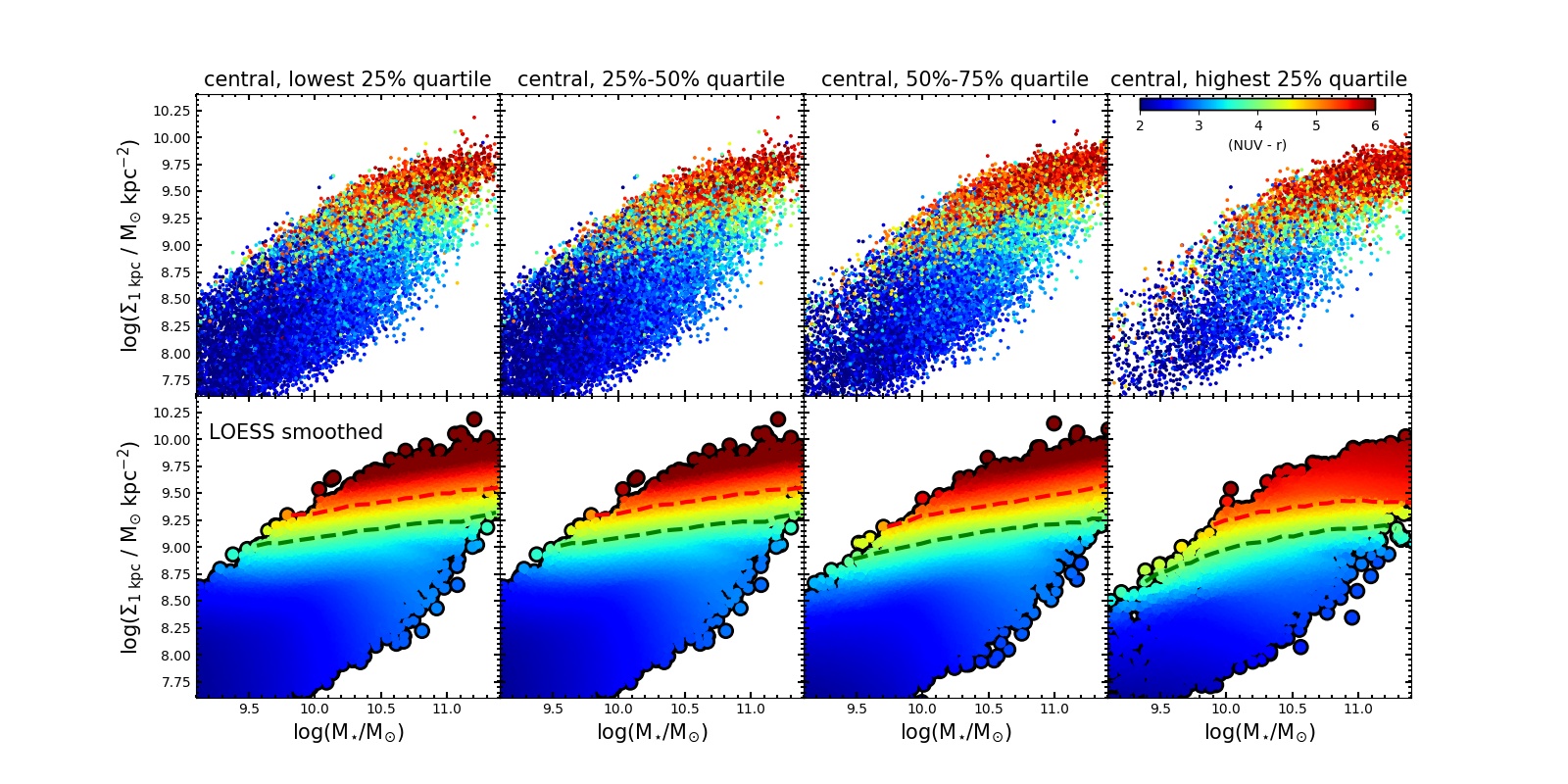}
   \caption{Central 1\ kpc surface mass density $\Sigma_{\rm 1\ kpc}$
        as a function of stellar mass for the central galaxies in four 
		environmental bins with increasing log($1+\delta$), color-coded by 
		the (NUV - r) color. The top panels show the scatter plots of the 
		original data; the bottom panels show the $V_{\rm max}$-weighted and
		LOESS-smoothed version. The green and red dashed lines denote the 
		transitional $\Sigma_{\rm 1\ kpc}$ in quenching with (NUV - r) $\sim$ 
		4 and in quenched status with (NUV - r) $\sim$ 5, respectively. 
              \label{fig:newcentral}}

\end{figure*}

In this appendix, we show the environmental impact on the color distribution
on the $\rm M_{\star}-\Sigma_{1\ kpc}$ plane for central galaxies. We present
the original data points color-coded by their (NUV - r) in the top panels of 
Figure \ref{fig:newcentral}; in the bottom panels, we repeat the same 
procedure of $V_{max}$-weighting and LOESS smoothing for the central 
galaxies as described in Section \ref{sec:ms+sigma1}. In the first two 
environment bins, both curves of $\rm \Sigma^{crit}_{1\ kpc}$ in quenching and 
quenched remain almost unchanged; as the overdensity increases, $\rm 
\Sigma^{crit}_{1\ kpc}$ in quenching at the low-mass end starts to decrease,
and the slope of the curve increases, which mimics the behavior of satellites. 
This could be due to the two-halo galactic conformity that a larger halo 
could cut off the gas supply to a surrounding smaller halo and hence quench 
the central galaxies inside the smaller halo \citep{lac21}. However, we 
caution that such similarity is more likely a manifestation of the 
misclassification of satellites as centrals due to the ``over-merging" of 
groups, which is a known issue for group-finder algorithms (e.g., see also
the top panel of Figure (1) in \citet{pen12}). Therefore, it is necessary 
to apply an additional cut on the overdensity to maximize the purity of the
``real" central galaxies. The insensitivity of $\rm \Sigma^{crit}_{1\ kpc}$
to the overdensity in the first two quartiles reconciles the validation of 
using the central galaxies at the lowest 25\% quartile as a representative 
sample of the ``real" central galaxies.

\section{The Parameterization of critical central density on stellar mass and overdensity} \label{appen:param}
\begin{figure*}
   \plotone{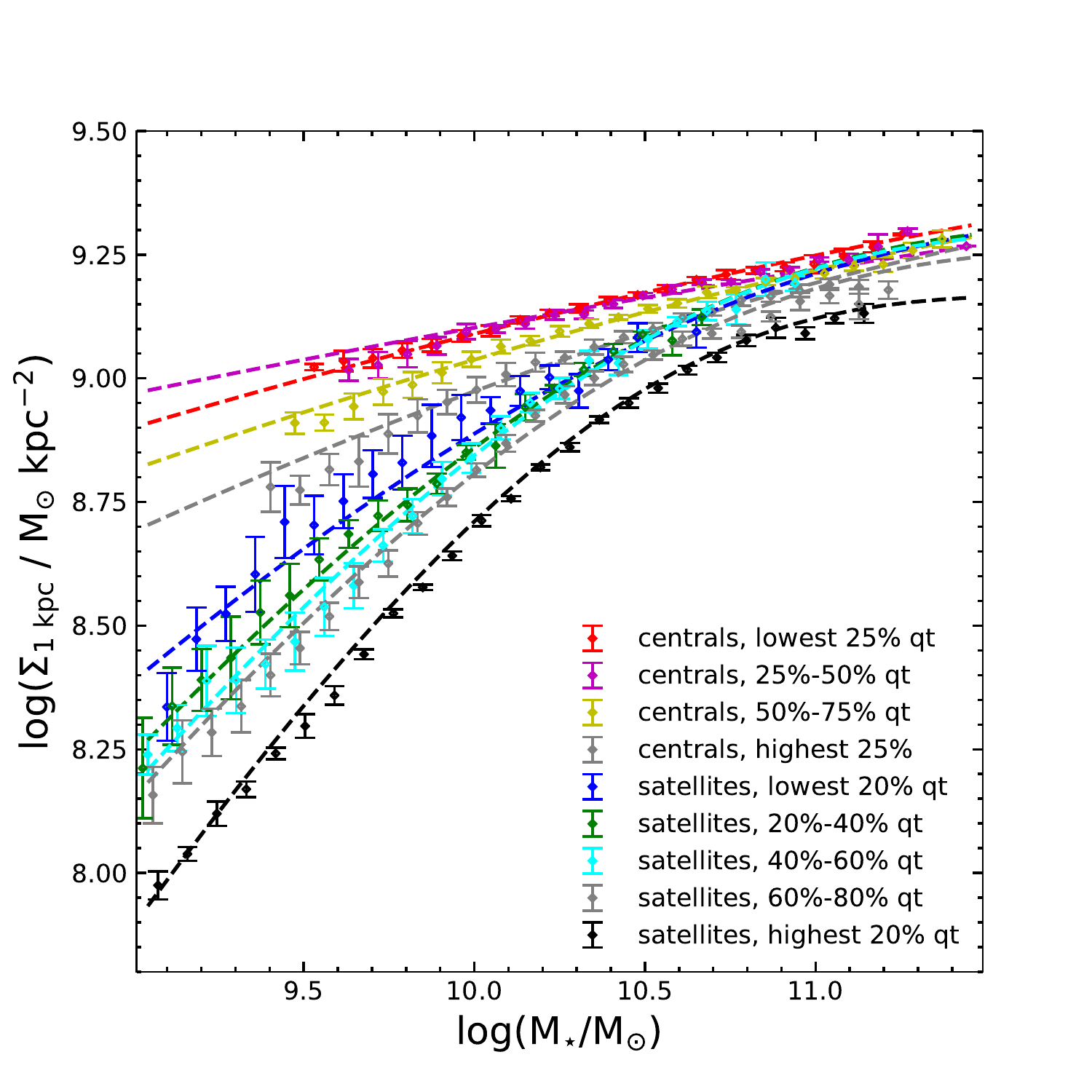}
   \caption{Transitional $\rm \Sigma^{crit}_{1\ kpc}$ in quenching (NUV -
		r $\sim$ 4) as a function of stellar mass for central (satellite) 
		galaxies in four (five) environment bins. The colored data points and 
		error bars denote the median $\rm \Sigma^{crit}_{1\ kpc}$ and 1$\sigma$
		error in each stellar mass bin computed over 30 jackknife resampling
		realizations for each environment bin, respectively. The colored 
		dashed lines are the best-fit lines which take the form of 
		Equation (\ref{eqn:form}) in each environment bin.
		\label{fig:param+fit}}	
\end{figure*}

\begin{figure*}
    \plottwo{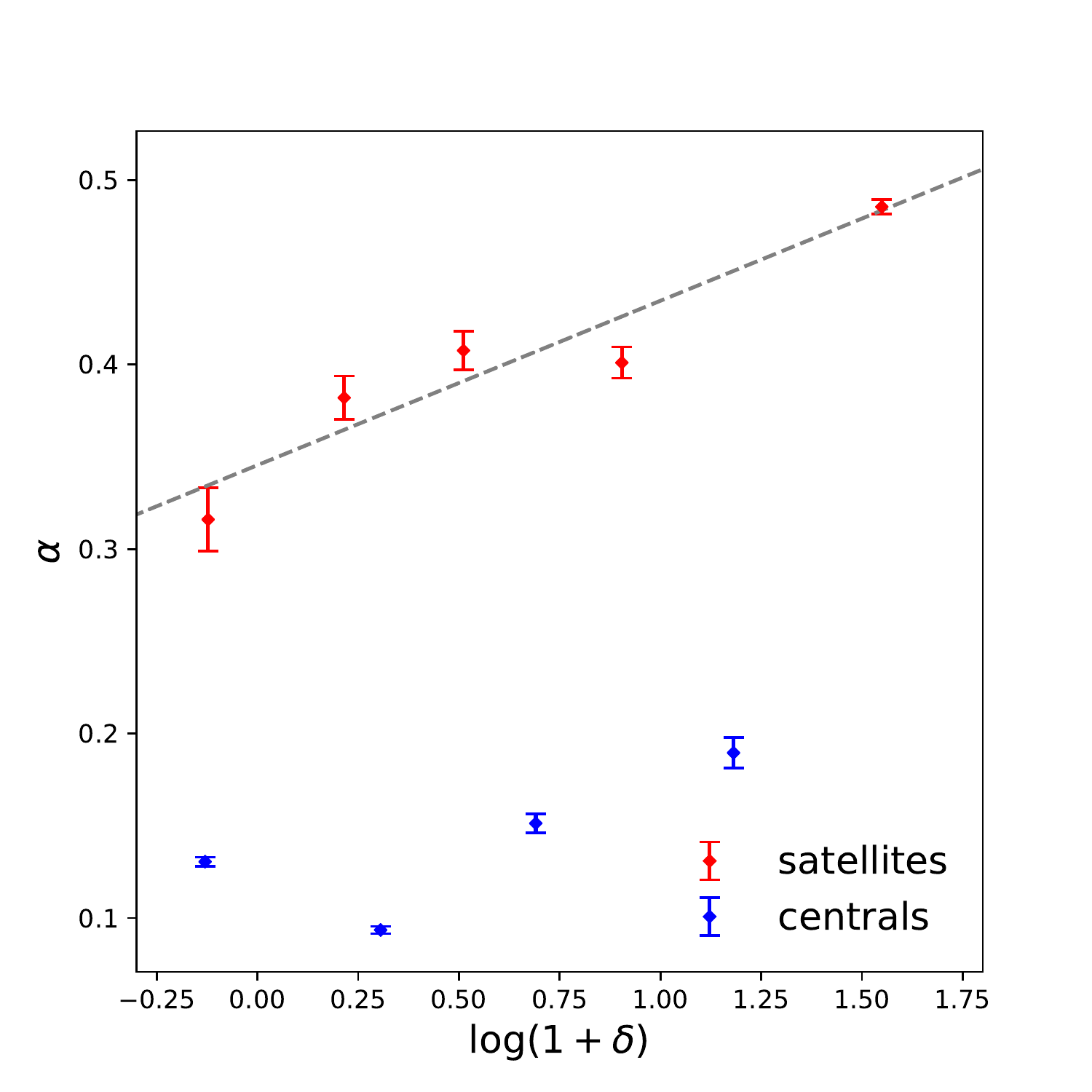}{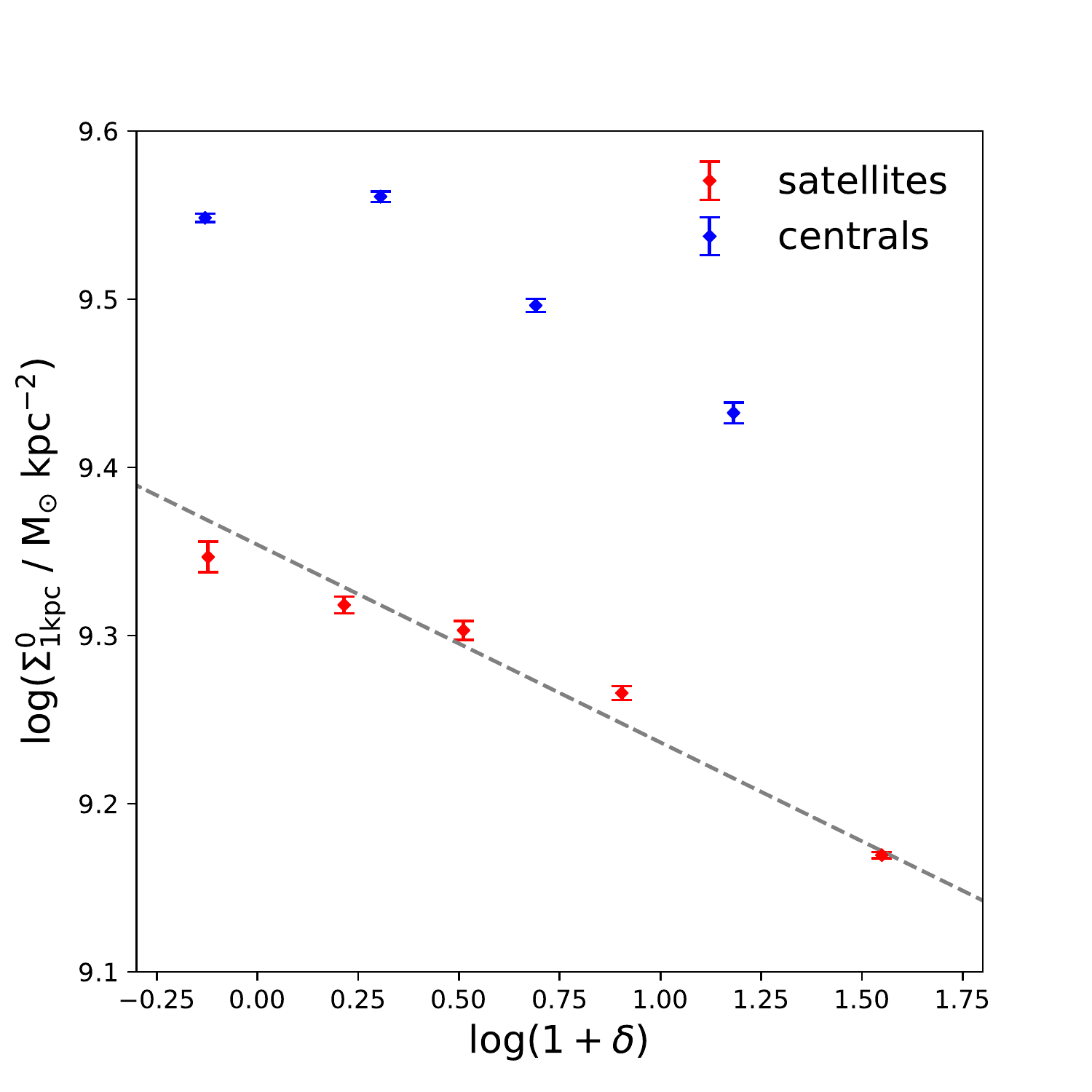}
		\caption{The left panel shows the slope $\alpha$ as a function of 
		log($1+\delta$); the right panel presents the normalization $\rm 
		\Sigma^0_{1\ kpc}$ as a function of log($1+\delta$). Blue (red) dots
		denote central (satellite) galaxies. The error bar denotes the 1$\sigma$
		error of the parameter from the fitting. The gray dashed lines are 
		best-fit lines of the correlations for satellites, which are 
		given in Equation (\ref{eqn:fit}).
         \label{fig:param+oden}}
 \end{figure*}

We show the parameterization of the dependence of $\rm \Sigma^{crit}_{1\ kpc}$
on the stellar mass and overdensity in this appendix. We adopt a similar
form of parameterization as that in \citet{zah17}, which is
\begin{equation}
    \rm \Sigma^{crit}_{1\ kpc}  = \Sigma^0_{1\ kpc}10^{ln(1 - exp(-(M_{\star}/M_b)^{\alpha}))}, 
\end{equation}
where $\rm \Sigma^0_{1\ kpc}$ is the normalization, $\alpha$ is the slope of 
the power law at the low-mass end, and $\rm M_b$ is the characteristic 
``bending" mass. In this study, we fix the bending mass as $\rm M_b = 
10^{10}M_{\odot}$. The physical motivation is that the bending mass marks
the transition of the mass-quenching process that is dominant in massive 
galaxies to the environment-quenching process that is dominant in low-mass 
galaxies. Observationally, the mass-quenching process typically proceeds in 
an inside-out mode that appears only to occur in galaxies with $\rm 
M_{\star} > 10^{10}M_{\odot}$  \citep{row18,blu20b}. To characterize its 
environmental dependence, we assign the satellite galaxies to five 
environment bins, and in each bin, we further divide the data into 25 
stellar mass bins. We perform the jackknife resampling in each environment 
bin 30 times and compute the median $\rm \Sigma^{crit}_{1\ kpc}$ and 1$\sigma$ 
error in each stellar mass bin. We present the best-fit curves in Figure 
\ref{fig:param+fit}, and provide the environmental dependence of the 
best-fit parameters for satellites in Figure \ref{fig:param+oden}. Since 
it is not clear whether the bending of the curves at higher overdensity 
for centrals is due to the real environmental effects or the 
misidentification of satellites as discussed in Appendix 
\ref{appen:central}, we do not characterize the environmental dependence for
the parameters of centrals.

The slope $\alpha$ of the satellites increases with the overdensity, and the 
normalization $\rm \Sigma^0_{1\ kpc}$ of the satellites decreases for 0.15 dex 
from lowest to highest density, whereas the centrals have systematically 
lower $\alpha$ and higher $\rm \Sigma^0_{1\ kpc}$. The dependence of 
$\alpha$ and $\rm \Sigma^0_{1\ kpc}$ on the overdensity for the satellites is
likely due to the combination of two things: (1) the star-forming level of
the satellites scales with the $\rm \Sigma_{1\ kpc}$ at fixed mass, as shown 
in the right panel of Figure \ref{fig:gthr}, so the quenched satellites typically 
have higher $\rm \Sigma_{1\ kpc}$ at fixed mass; and (2) the fraction of 
quenched satellites at fixed mass increases with the overdensity. See the detailed discussion in Section \ref{sec:dis}

\bibliography{}
\bibliographystyle{apj}

\end{document}